\documentstyle[epsfig,psfig]{texas}

\begin{document}

\title{INTENSITY CORRELATION BETWEEN OBSERVATIONS AT DIFFERENT WAVELENGTHS FOR Mkn 501 IN 1997}
\author{C\'ecile RENAULT$^{1}$  for the CAT collaboration, 
Nicolas RENAUD$^{2}$ and Gilles HENRI$^{2}$}
\address{(1)LPNHE Paris VI et VII \\
4 place Jussieu, Tour 33 - Rez de chaussée, F-75252 Paris Cedex 05 \\
}
\address{(2)Observatoire de Grenoble \\
BP 53, F-38041 GRENOBLE Cédex 9\\
{\rm Email: rcecile@in2p3.fr, nrenaud@obs.ujf-grenoble.fr, Gilles.Henri@obs.ujf-grenoble.fr }}

\begin{abstract}
The CAT imaging telescope on the site of the former solar plant Th\'emis
 in southern France  observed $\gamma$-rays 
from the BL~Lac object Mkn501 above 250~GeV for more than 60~usable hours  
on-source from March  to October~1997. This source was in a state 
of high activity during all this period. By studying the correlation 
between the photons of different energies detected by the CAT imaging telescope 
and by the ASM/RXTE experiment (1.3-12.0~keV) on board the {\it Rossi X-Ray Timing
Explorer}, we may constrain the mechanisms which could lead to 
the emission of these photons.
\end{abstract}

\section{Introduction}
During several months in 1997, the Active Galactic Nuclei Mkn~501 (z=0.034)
 went into a very high state of activity. 
It is the second closest BL~Lac object and one of the two extra-galactic sources confirmed 
at very high energy (the other being Mkn~421).
The source was extensively observed at wavelengths ranging from radio to VHE (Very High Energy)
gamma-rays.  Here we use essentially X-ray and gamma-ray information respectively from ASM (All Sky Monitor) on 
board RXTE and CAT (Cherenkov Array at Th\'emis). 

The spectral energy distribution of Mkn~501 exhibits two bumps. In agreement with the unified scheme
of the AGN (\cite{urry}), the first peak with a maximum at 10-100 keV (\cite{pian})
is thought to be produced by synchrotron emission of particles in a jet pointing towards us; the second peak 
culminates around 1 TeV (\cite{aharon},\cite{cat}), making Mkn~501 the hardest BL~Lac object
ever observed.
In Section 2, the data sample is described while 
Section 3 provides information about theoretical models. Then, Sections 4 and 5 are respectively 
devoted to the search for variability and micro-variability in the VHE band. Sections 6 and 7
presents studies of correlations between emissions in X-ray and gamma-ray
and correlation between different bands inside the VHE domain.

\section{Data sample}
The 17.8~m$^2$ CAT  imaging telescope started operation on the site of the former
 solar plant Th\'emis in the French Pyr\'en\'ees (southern France) in Autumn 1996. A very high definition
camera of 600~phototudes (4.8$^\circ$ field of view) allows an analysis using the Cherenkov
light distribution inside the image of the cosmic-ray shower. A complete description
 of the telescope and the 
camera can be found in \cite{barrauNIM}, \cite{barrau} and details about the 
analysis method are available in  \cite{lebohecNIM}, \cite{lebohec}. 

After data cleaning, a total of 57.2 hours of observation of Mkn~501 and 22.5 hours on control regions 
is used to compute the light curve above 300~GeV.

This BL~Lac object is also monitored on a regular basis
by the All Sky Monitor (ASM) on board the {\it Rossi X-Ray Timing
Explorer} (RXTE) (\cite{asm}), providing information  about the
X-ray activity in the energy region from 2 to 12 keV.  The ASM count rates are determined from the 
``definitive'' ASM data which have a dwell duration larger than 30~seconds 
and a flux fit with a reduced $\chi^2$-value below 1.5. 
The light curve is extracted using the ``ftools 4.0'' package.

\begin{figure}
\centering
\psfig{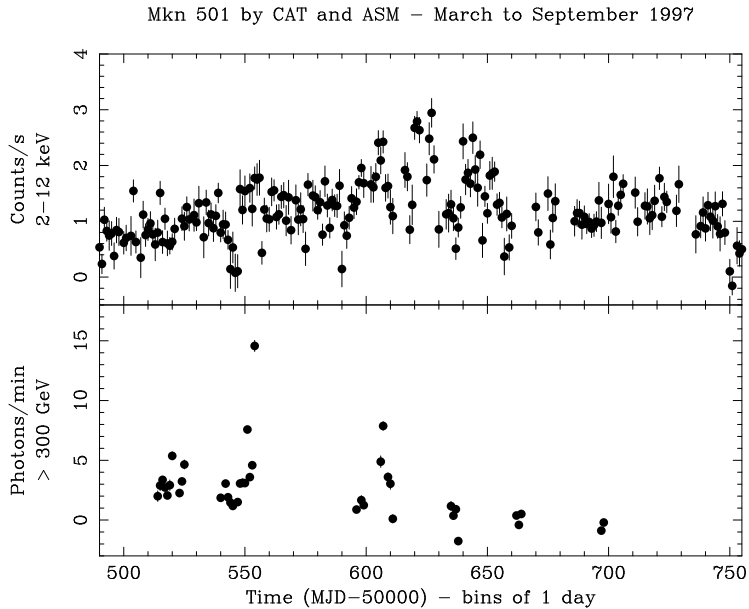}
\caption{Light curve of Mkn 501 from March to September 1997 with a time binning of 1 day. Upper panel:
X-ray data from ASM, lower panel: gamma-ray data from CAT. }\label{lc}
\end{figure}

Figure \ref{lc} presents the light curve obtained with the CAT and ASM data 
between March and September 1997.

\section{Models}

In the case of leptonic model, the high energy emission of blazars can be 
well explained by Inverse Compton
emission of relativistic particules. As BL~Lacs objects are characterized 
by the weakness of their thermal component and of their emission lines, the 
principal source of photons for the Compton interaction should be the 
synchrotron emission radiated by the high energy particles (SSC model).
In the case of hadronic models the high energy emission is produced by 
pair cascades resulting from initial photopion processes.    
In the both cases observation of photons with an energy of $\gamma m_e c^2$
requires particles with a Lorentz factor at least $\gamma$.
Particle acceleration is so an important ingredient of all model with emission at high 
energy. The variability and correlation between different wavelengths of this emission can 
give severe constraints on both models.

\section{Search for short-term variability}

The search for short term variability is of special importance as it provides immediately an upper limit
on the source size $R$ for a given Doppler factor $\delta$ with few assumptions needed. 
In order to confront a specific model or define a region in the ($R$,$\delta$) plane, 
few hypothesis must be made.

While variability at a daily scale is directly seen Fig. \ref{lc}, the search for intra-night variability
must  be studied in more detail. In this paper the aim is not a systematic study
of all scales of variability but the identification of a rise time, {\it i.e.} the necessary time
for doubling the flux with a significance of at least 3 $\sigma$.
 However, as the source is observed for only about one-two hours per night,
 we can not check sub-day variations with durations larger about 30 minutes.
At the other extreme, if we are interested in a significant rise time of 1-10 minutes, the flux has to be larger 
than 6 gamma per minute. Only three nights satisfied this requirement: MJD 50551.08 (8.7 $\gamma$/min), 
 50554.13 (14.3 $\gamma$/min) and 50606.96 (7.7 $\gamma$/min).

\begin{figure}
\centering
\psfig{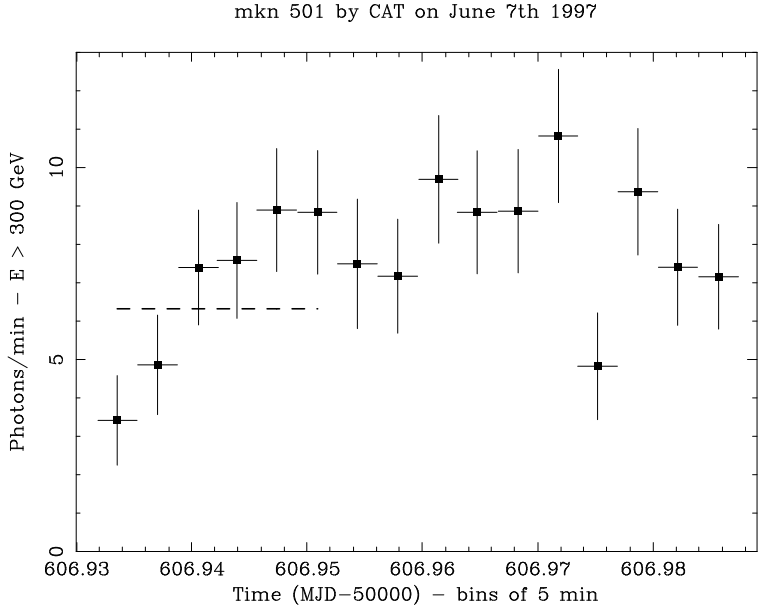}
\caption{Light curve of Mkn 501 for June 7 with a threshold of 300~GeV.}\label{rise}
\end{figure}

The night of April 16, with the strongest flare (MJD 50554), appears to be flat when it is studied with 
time binnings of 1 to 10 minutes; this is also true for
 the night of April 13. Only  June 7 exhibits a lightcurve
with a regular increase of the flux at the beginning of the night (from 3.5 to 9 $\gamma$/min) in 
 $\Delta t \approx$30 minutes, as shown in Fig. \ref{rise}. The $\chi^2$ of the fit by a constant on the first 6 
points is 13.7.

Thus  the size $R$ of the emission region must be less than $c \delta \Delta t$ which leads to
$R < 5.4\ 10^{13} \times \delta$~cm. It is possible to combine this limit with two other constraints:
\begin{itemize}
\item{$\gamma-\gamma$ opacity $<$ 1, to allow TeV photons to escape (upper limit)};
\item{ compatibility with the observed ratio Lum$_{sync}$/Lum$_{IC}$ from \cite{pian} (lower limit)}.
\end{itemize}

\begin{figure}
\centering
\psfig{file=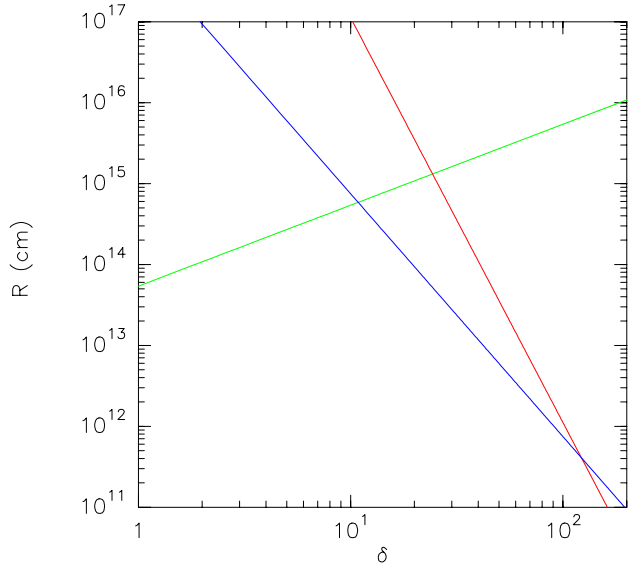,width=8cm,angle=-90}
\caption{Permitted values in the plane ($R$,$\delta$). The upper line (green) shows the upper limit put by the
rise time of 30~minutes. The left line (blue) 
indicates the upper limit induced by the requirement that  $\gamma-\gamma$ opacity 
be lower than 1. The right line (red) indicates the lower limit  dictated by the compatibility with the
 observed ratio Lum$_{sync}$/Lum$_{IC}$ from \cite{pian}.
}\label{Rd}
\end{figure}

Fig. \ref{Rd} presents the results obtained assuming an homogeneous source
and a particle distribution in $\gamma^2 \ exp(-\gamma / \gamma_{max})$ 
which leave a very restricted domain with $R< 10^{12}$~cm and
$\delta >$ 100. These values are quite unrealistic, the Doppler factor being too high. So the hypothesis 
of homogeneity should be certainly revised for further studies. Moreover, in such a homogeneous model, 
we would expect a very strong  short-term correlation between X-ray and gamma-ray which is not observed
(see Section 6).

\section{Search for micro-variability}

\begin{figure}
\centering
\psfig{file=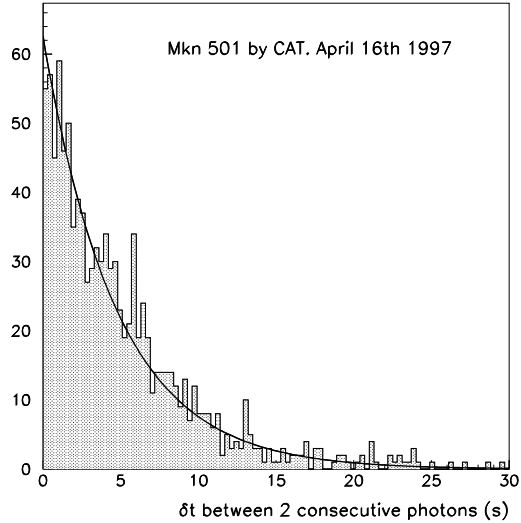,width=8cm}
\caption{Distribution of the time difference between two consecutive photons  in seconds.
The distribution is well fitted  by an exponential law (full line).}\label{dt}
\end{figure}

As suggested by M. Urry (private communication), a flare could result of the superimposition of many 
``micro-flares''
with duration of a few seconds. We can study the time arrival distribution of the photons for the 1.5 hours
of observation taken the April 16$^{th}$. Because of the very high flux ($\approx$ 8$\times$ the Crab flux),
 we can directly use the ``ON'' data which contain about 90\% of gammas. The result is presented
in Fig. \ref{dt}. No deviation from the expected exponential distribution is observed: no flares of
a few seconds contribute significantly to the very high flux observed during this night.

\section{Search for correlation X-gamma}
\begin{figure}
\centering
\psfig{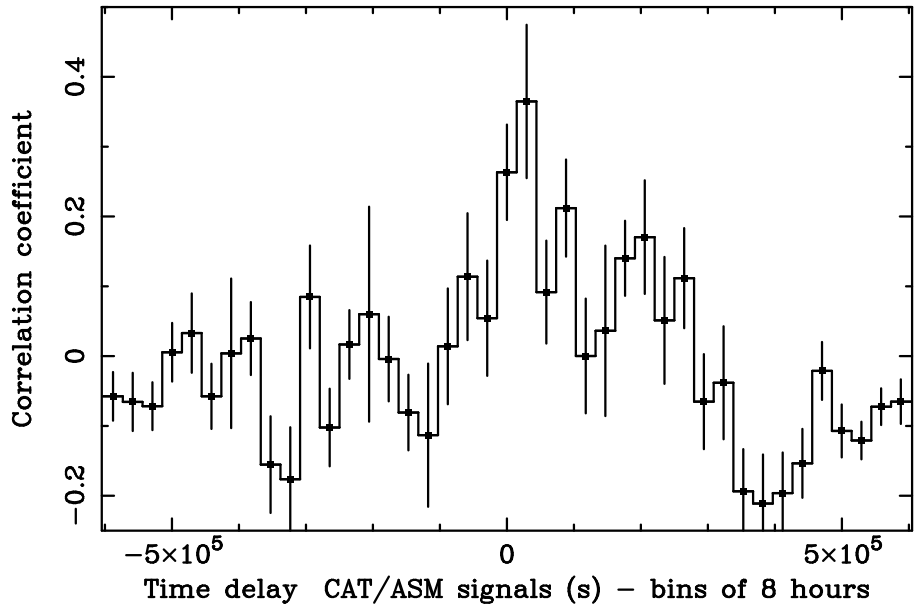}
\caption{Correlation coefficient in function of the time delay between X-ray radiation (ASM, 2-12 keV)
and gamma-ray
emission (CAT, $>$ 300~GeV). A positive delay means that variations in the TeV range lag those in  the
X-ray band. It is computed from data of April and June  1997 (except 16$^{th}$).}\label{corxg}
\end{figure}

A correlation between X-ray emission,  due to synchrotron radiation and gamma-ray emission, assumed
to be produced by inverse-Compton scattering would reinforce the fact that the same population of particles
is at the origin of both emissions. It does not presume of the nature of these particles, leptonics or 
hadronics. Fig \ref{corxg} presents the nightly correlation of the flux of Mkn~501 as measured by ASM and CAT.
We restricted the sample to the  months April and June when the flux was variable and enough data were 
available. The night April 16 is not included in the sample because it strongly dominates the fluctuations
in the  TeV band and does not correspond to any flare in the 2-12 keV band. Unfortunately,
it is not possible to study correlation at shorter time scales because of the differences in time sampling:
CAT observe the source consecutively for 1 to 3 hours per day
while ASM takes data for 90 seconds more or less every hour: the statistics in the X-ray band
is not significant per bin of a few hours only.

\begin{figure}
\centering
\psfig{file=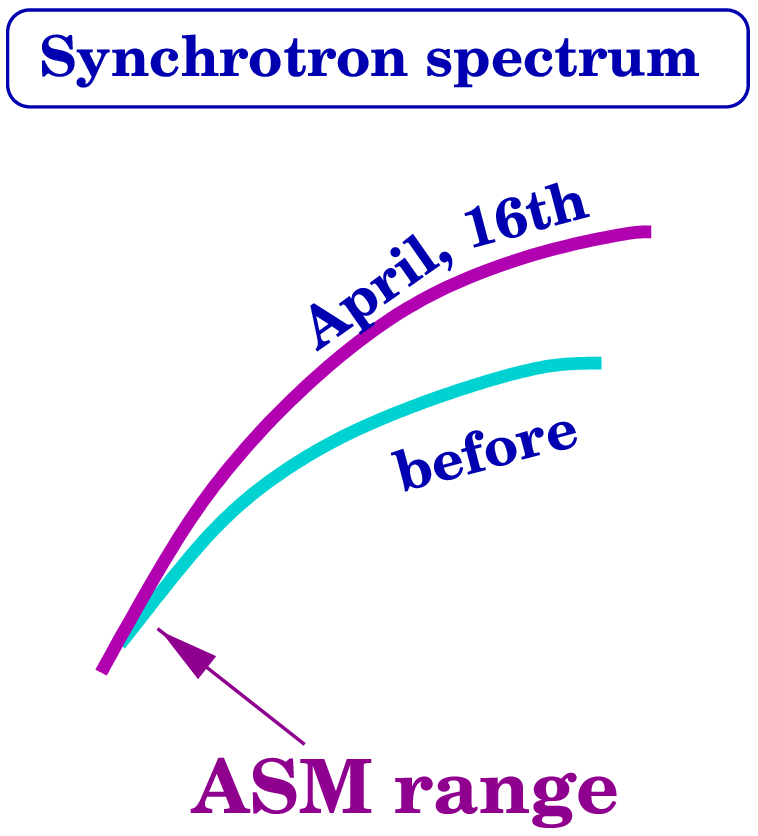,width=6cm}
\caption{Schema of the evolution of spectral
energy density explaining why the flare of April 16 is not observed by ASM.}\label{sed}
\end{figure}

The quite low correlation coefficient ($\approx$0.35$\pm$0.10) could be explained by the differences
in time sampling. If we can not
put by this way constraints on delays between X-ray and gamma-ray fluctuations, we have evidence of simultaneous
evolution in both energy ranges, comforting the hypothesis of a common origin for both emissions.
The lack of ASM flare corresponding to the April 16
 gamma flare, despite an increase of the flux observed by BeppoSAX
(\cite{pian}) in the 1-200~keV band simultaneously with the CAT flare, 
is understandable if the evolution of the spectral
energy density follow the evolution schematized in Fig. \ref{sed}.

\section{Search for correlation gamma-gamma}

\begin{figure}
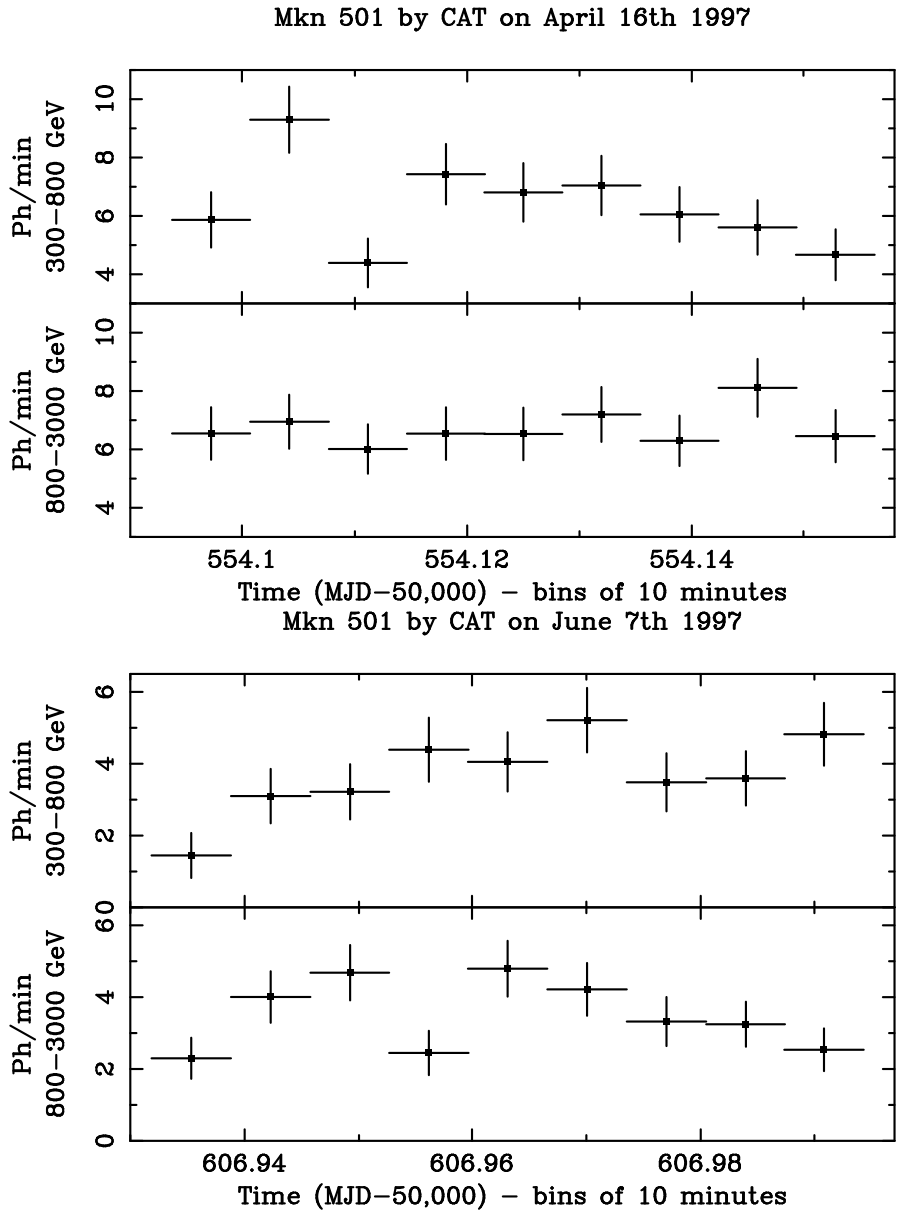

\centering
\psfig{file=01_CRENAULT1_7.ps,width=8cm,angle=-90}
\psfig{file=01_CRENAULT1_8.ps,width=8cm,angle=-90}
\caption{Lightcurve of Mkn~501 observed by CAT in 1997 in the ranges 300-800~GeV and 800-3000~GeV
with time binning of 10 minutes. Upper panel: a flat night (April 16) and lower panel: the night June 7
with a  rising time of 30~minutes at the beginning.}\label{corgg}
\end{figure}

One can also search for time delay between ``hard'' and ``soft'' CAT photons. In an
inhomogeneous model with pair creation (as described in \cite{alex} for an external model), the most energetic photons 
should lag the softer ones because they can escape later. In order to test this hypothesis, the
lightcurves in two energy bands were computed for two nights (see Fig. \ref{corgg}). Whatever the total 
flux is constant or increasing, no  experimental evidence for time delay 
is seen. 
This chaotic behaviour needs a detail study of a time-dependent 
inhomogeneous model. Such models are in progress (for a model only including 
external Compton interaction see Renaud \& Henri, these proceedings).  

\section{Conclusion}

During its strong outburst of 1997, the BL~Lac object Mkn~501 was extensively observed at many wavelengths,
in particular  by All Sky Monitor in the X-ray band and  by CAT at TeV energies. The search for variability
and correlations can give clues for the understanding of the still most mysterious class of AGN.
With a rise time of 30~minutes and a nightly correlation between X-ray and gamma-ray emissions, 
 models with a same population of particles, leptonic or hadronic, yielding in a compact zone synchrotron
radiation and inverse-Compton emission are reinforced. More detailed studies indicates that a simple
homogeneous model can not account for the observations and further refinements like inhomogeneity 
seem to be necessary.


\end{document}